\RequirePackage{etex}
\RequirePackage{graphicx}

\documentclass{iau}

\usepackage{amsmath}
\usepackage{graphicx}
\usepackage{multirow}

\usepackage{siunitx}
\usepackage[capitalize]{cleveref}

\usepackage{./aasmacros}
\DeclareSIUnit\parsec{pc}
\DeclareSIUnit\year{yr}
\DeclareSIUnit\years{yrs}
\DeclareSIUnit\dex{dex}
\DeclareSIUnit\mag{mag}
\DeclareSIUnit\h{h}
\DeclareSIUnit\Msun{M_\odot}
\DeclareSIUnit\Rsun{R_\odot}
\DeclareSIUnit\Lsun{L_\odot}

\newcommand*{\OIII}{[O\,\textsc{iii}]} 

\begin{document}

\lefttitle{Lucas M. Valenzuela et al.}
\righttitle{PICS: PNe In Cosmological Simulations}

\journaltitle{Planetary Nebulae: a Universal Toolbox in the Era of Precision Astrophysics}
\jnlDoiYr{2023}
\doival{10.1017/xxxxx}
\volno{384}

\aopheadtitle{Proceedings IAU Symposium}
\editors{O. De Marco, A. Zijlstra, R. Szczerba, eds.}
 
\title{PICS: Planetary Nebulae in Cosmological Simulations -- Revelations of the Planetary Nebula Luminosity Function from Realistic Stellar Populations}

\author{Lucas M. Valenzuela$^1$, Rhea-Silvia Remus$^1$, Marcelo M. Miller Bertolami$^{2,3}$, and Roberto H. Méndez$^4$}
\affiliation{$^1$Universitäts-Sternwarte, Fakultät für Physik, Ludwig-Maximilians-Universität München, Scheinerstr. 1, 81679 München, Germany}
\affiliation{$^2$Instituto de Astrofísica de La Plata, UNLP-CONICET, La Plata, Paseo del Bosque s/n, B1900FWA, Argentina}
\affiliation{$^3$Facultad de Ciencias Astronómicas y Geofísicas, UNLP, La Plata, Paseo del Bosque s/n, B1900FWA, Argentina}
\affiliation{$^4$Institute for Astronomy, University of Hawaii, 2680 Woodlawn Drive, Honolulu, HI 96822, USA}

\begin{abstract}
Even after decades of usage as an extragalactic standard candle, the universal bright end of the planetary nebula luminosity function (PNLF) still lacks a solid theoretical explanation.
Until now, models have modeled planetary nebulae (PNe) from artificial stellar populations, without an underlying cosmological star formation history.
We present PICS (PNe In Cosmological Simulations), a novel method of modeling PNe in cosmological simulations, through which PN populations for the first time naturally occur within galaxies of diverse evolutionary pathways.
We find that only by using realistic stellar populations and their metallicities is it possible to reproduce the bright end of the PNLF for all galaxy types. In particular, the dependence of stellar lifetimes on metallicity has to be accounted for to produce bright PNe in metal-rich populations.
Finally, PICS reproduces the statistically complete part of the PNLF observed around the Sun, down to six orders of magnitude below the bright end.
\end{abstract}


\maketitle

\section{Introduction}

As bright sources of the nebular emission line \OIII{} $\lambda5007$, planetary nebulae (PNe) have long been used as tracer populations and distance indicators in other galaxies, meanwhile reaching distances as far away as \SI{40}{\mega\parsec} \citep{jacoby+23}. While the bright end of their luminosity function (planetary nebula luminosity function, PNLF) has observationally been found to be universal among different types of galaxies and thus letting PNe be used as standard candles \citep[e.g.,][]{ciardullo+89:pnlfII,jacoby89:pnlfI}, the theoretical background of the PNLF is still not well understood \citep[e.g.,][]{ciardullo22}. For that reason, numerous models have attempted to improve our understanding of the bright end of the PNLF \citep[e.g.,][]{dopita+92,gesicki+18,souropanis+23,yao&quataert23} and its overall shape \citep[e.g.,][]{mendez&soffner97,valenzuela+19,chase+23}. These models have generally been based on artificial stellar populations, oftentimes using the results from studies of Milky Way stars. However, such stellar populations may not be representative of the diverse star formation histories that lead to a wide range of stellar populations in other galaxies, for example in old, metal-rich elliptical galaxies \citep[e.g.,][]{li+18:gradients}.

With the advent of hydrodynamical cosmological simulations in the past decade, it has become possible to obtain more realistic stellar populations with a large variety of cosmological star formation histories that are self-consistent with the actual assembly history of the simulated galaxy itself. In this work, we introduce the PNe In Cosmological Simulations (PICS) method, which for the first time can use simulated stellar populations as the underlying systems for PN populations and the PNLF. We present the PICS method and the models used in \cref{sec:model}. In \cref{sec:results}, we present first results from applying PICS to a cosmological simulation, focusing on the relevance of metallicity and the reproduction of the Milky Way PNLF. Finally, we summarize and conclude our findings in \cref{sec:conclusion}.

\section{Model}
\label{sec:model}

To introduce PNe in cosmological simulations, the Planetary nebulae In Cosmological Simulations (PICS) method provides a modular framework for determining the population of PNe for a given single stellar population (SSP). As each stellar particle in a cosmological simulation represents an SSP, this allows one to apply PICS to all the stellar particles in such a simulation in post-processing. As a result, PNe are obtained with all the inherited properties from their parent stellar particles, such as spatial and kinematic information or their ages, and with the properties of the PNe themselves, such as the \OIII{} $\lambda5007$ (from hereon simply referred to as \OIII{}) magnitude or the central star mass.

The framework consists of a sequence of models that determine the PNe population for an SSP, which is defined by its age, total mass, initial mass function (IMF), and its element abundances. In its basic form, the sequence is the following:
\begin{enumerate}
    \item Lifetime function for the SSP, through which the initial mass of the stars reaching the post-asymptotic giant branch (AGB) phase can be determined from the SSP age.
    \item Initial-to-final mass relation (IFMR), from which the central star mass is obtained.
    \item Post-AGB stellar evolution tracks, which provide the central star properties for a given post-AGB age, that is the time passed since leaving the AGB. In particular, the stellar properties consist of the effective temperature and luminosity of the star.
    \item Planetary nebula model, which determines the nebular properties, most importantly the \OIII{} magnitude.
\end{enumerate}

\begin{figure}
    \begin{center}
    \includegraphics[width=0.49\textwidth]{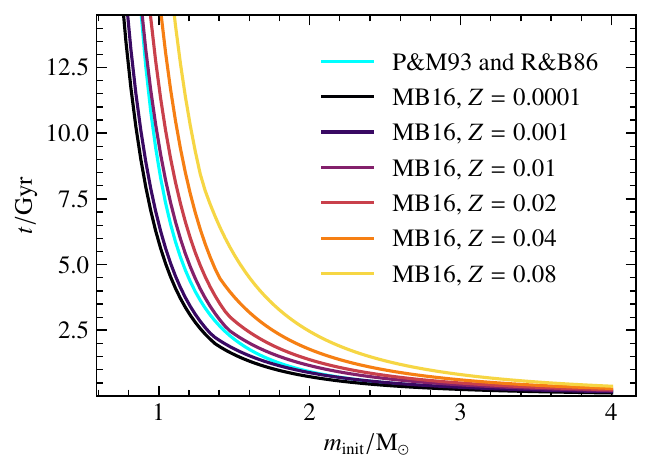}
    \includegraphics[width=0.49\textwidth]{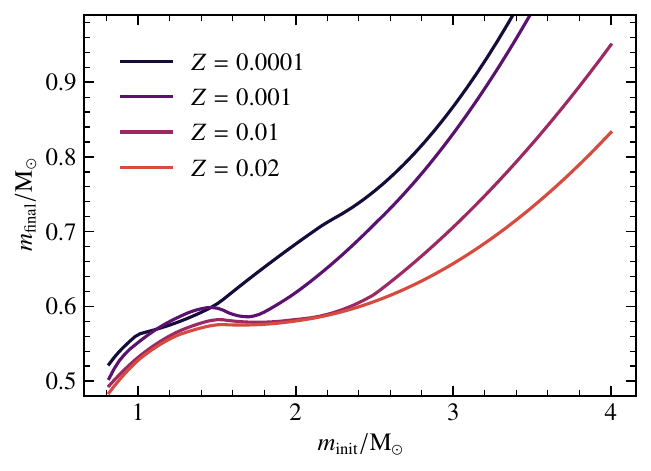}
    \end{center}
    \caption{
        \emph{Left}: Lifetime functions from the main sequence plus post-main sequence lifetime functions by \citet{padovani&matteucci93} and \citet{renzini&buzzoni86}, respectively, and from the total lifetimes as computed by \citet{miller_bertolami16} for different metallicities, $Z$, given the initial mass of a star. The lifetime functions for $Z=0.04$ and 0.08 are extrapolated from the four computed ones.
        \emph{Right}: Initial-to-final mass relations for the four computed metallicities from \citet{miller_bertolami16}.
    }
    \label{fig:lifetime_ifmr} 
\end{figure}

For (1), the lifetime function, we consider two different prescriptions: The first is composed of the main sequence lifetime from \citet{padovani&matteucci93}, which is the one used in the Magneticum Pathfinder simulations \citep[the implemented stellar evolution models are described by][]{tornatore+07}, and the post-main sequence lifetime from \citet{renzini&buzzoni86}. The second is the group of lifetime functions as obtained from \citet{miller_bertolami16}, which vary with metallicity. The left panel of \cref{fig:lifetime_ifmr} shows the lifetime functions, where it can be seen that the lifetime function from \citet{padovani&matteucci93} and \citet{renzini&buzzoni86} is most similar to the lifetime function at a metallicity of $Z=0.01$ from \citet{miller_bertolami16}. It also becomes apparent that more metal-rich stars reach the PN phase after significantly longer times than metal-poor stars of the same initial mass. This is due to the impact of metallicity on bound-free opacities, which leads to larger luminosities at lower metallicities.

For (2) and (3), the IFMR and post-AGB stellar evolution tracks, respectively, we use the metallicity-dependent results from \citet{miller_bertolami16}. The IFMRs for the four metallicities used by \citet{miller_bertolami16} are shown in the right panel of \cref{fig:lifetime_ifmr}. All of the models are interpolated between the metallicities. The lifetime functions are extrapolated beyond the originally computed metallicities, as running new stellar evolutionary simulations for the higher metallicities is beyond the scope of this work. In future studies, we will have such simulated results available to replace the extrapolations. Finally, for (4), the PN model, we use the empirical model presented by \citet{valenzuela+19}, which is also based on the assumption that PNe can be optically thin, thus losing some of their \OIII{} intensity.
The model will be presented in detail in a full paper (Valenzuela et al.\ in prep.).

\section{Application to a Cosmological Simulation}
\label{sec:results}

As a first test in applying the model to a cosmological simulation, we use the Magneticum Pathfinder\footnote{\url{www.magneticum.org}} Box4 (uhr) simulation, which has a side length of \SI{68}{\mega\parsec} and an average stellar particle mass of \SI{1.8e6}{\Msun}. We assume a Chabrier IMF \citep{chabrier03:imf} as also used within the simulation. For more details on the simulation and an overview of comparisons with observations, see \citet{teklu+15} and \citet{valenzuela&remus22}, respectively.

\subsection{The Importance of Metallicity}

In recent studies of the PNLF through modeling, it has been customary to only consider PNe of a single fixed solar-like metallicity of $Z=0.01$ or $Z=0.02$ \citep[e.g.,][]{gesicki+18,valenzuela+19,souropanis+23,yao&quataert23}. To evaluate the importance of taking the actual metallicities of PNe into account, we apply PICS with two different model sequences to the cosmological simulation. For the first, we only use the relations given for a metallicity of $Z=0.01$ from \citet{miller_bertolami16} and the lifetime function from \citet{padovani&matteucci93} and \citet{renzini&buzzoni86}. For the second, we apply the full metallicity-dependent models to the stellar particles.

\begin{figure}
    \begin{center}
    \includegraphics[width=\textwidth]{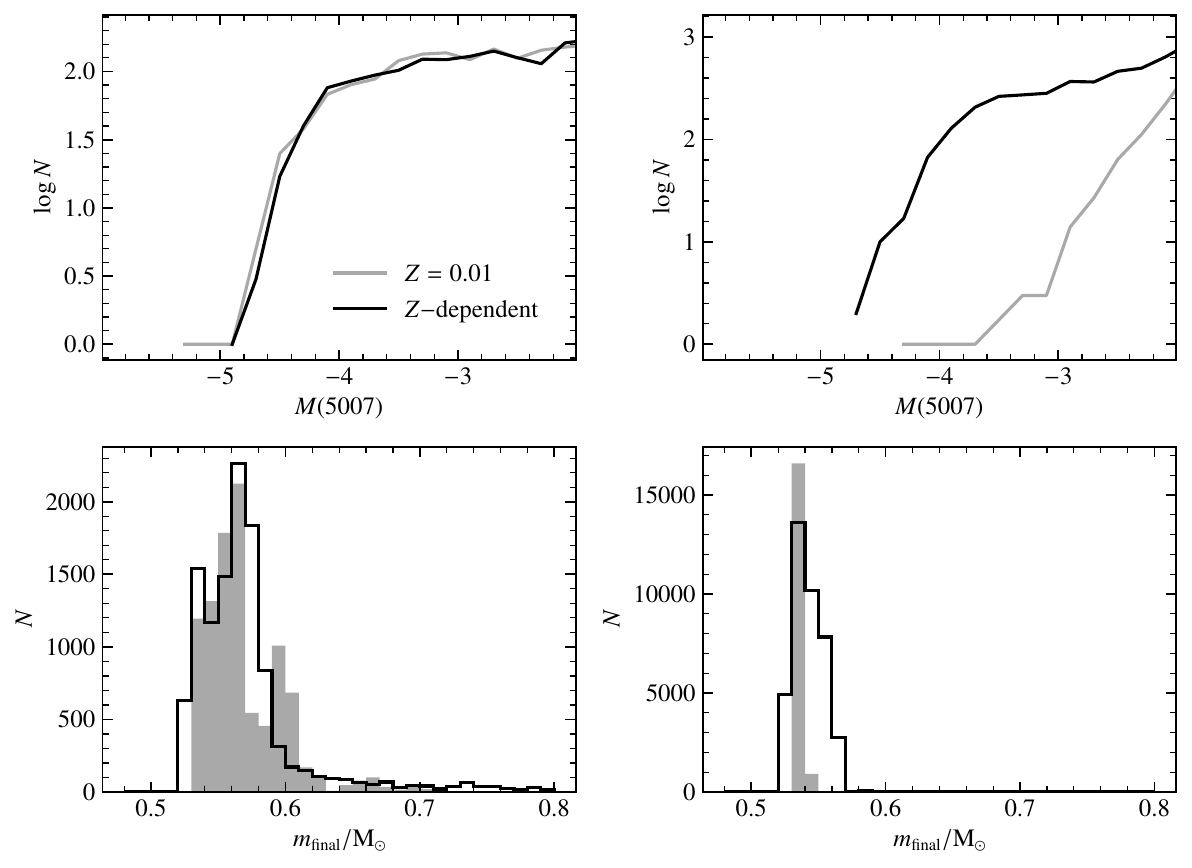}
    \end{center}
    \caption{\emph{Top}: Obtained PNLFs for two simulated galaxies (left and right panel each show the results for one galaxy) by running PICS with the models for $Z=0.01$ (gray) and with the full metallicity-dependent models (black). \emph{Bottom}: Final mass distributions of the PN central stars for the same two galaxies and running the same two varieties of PICS model sequences (in gray and black as for the top row).}
    \label{fig:pnlfs_mfinal_met} 
\end{figure}

We applied the two model sequences of PICS to two different galaxies in the simulation, thus obtaining two PNe populations with \OIII{} magnitudes for each galaxy, leading to two PNLFs per galaxy. This led to very different results between the two galaxies. The first galaxy, which has a stellar mass of $M_* = \SI{1.8e10}{\Msun}$, an average stellar age of $\langle t_\mathrm{age} \rangle = \SI{6.5}{\giga\year}$, and an average metallicity of $\langle Z \rangle =0.013$, features the well-known PNLF bright end cutoff for both model sequences (top left panel of \cref{fig:pnlfs_mfinal_met}). In contrast, the second galaxy, which is much more massive, older, and more metal-rich ($M_* = \SI{1.2e11}{\Msun}$, $\langle t_\mathrm{age} \rangle = \SI{9.3}{\giga\year}$, $\langle Z \rangle = 0.022$), only features a normal PNLF for the metallicity-dependent model sequence, but completely lacks the brightest PNe when disregarding the metallicity on the stellar lifetimes (top right panel of \cref{fig:pnlfs_mfinal_met}).

The reason for the large differences between the two PNLFs of the second galaxy can be found in the distribution of the final masses of the PN central stars (bottom row of \cref{fig:pnlfs_mfinal_met}). While the $Z=0.01$ model predicts final stellar masses of mostly $M_\mathrm{final} < \SI{0.54}{\Msun}$, the metallicity-dependent model finds final masses of up to $M_\mathrm{final,max} = \SI{0.57}{\Msun}$, which is only slightly lower than the required maximum final mass as obtained by \citet{valenzuela+19} for old stellar populations. This is a consequence of the lifetimes being longer for any given stellar mass at higher metallicities (left panel of \cref{fig:lifetime_ifmr}), leading to the metal-rich, more massive stars reaching the PN phase at later times than metal-poor stars of the same mass. As the second galaxy is old and relatively metal-rich, the effect is strongly visible, whereas the first galaxy has metallicities close to the assumed $Z=0.01$ of the metallicity-independent model, thus only leading to minor changes in the PNLF and final mass distribution. As a result, the metallicities clearly cannot be neglected in the general case when modeling the PNLF.

\subsection{Normalization and Shape of the PNLF}

\begin{figure}
    \begin{center}
    \includegraphics[width=0.5\textwidth]{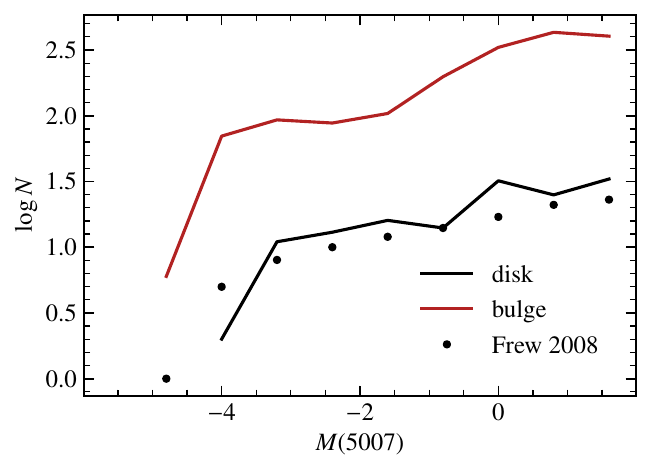}
    \end{center}
    \caption{Planetary nebula luminosity functions of PNe in the Milky Way within \SI{2}{\kilo\parsec} of the Sun from \citet{frew08} (black dots), for the modeled PNe within a sphere of \SI{2}{\kilo\parsec} radius with its center in the disk, \SI{8}{\kilo\parsec} away from the center of a selected simulated galaxy (black line), and for the modeled PNe within \SI{2}{\kilo\parsec} of the center of the simulated galaxy (red line). The simulated galaxy is a disk galaxy with a mass similar to that of the Milky Way. The PNe were modeled using the metallicity-dependent lifetime functions. Note that the $M(5007)$-axis reaches much deeper magnitudes than the ones in \cref{fig:pnlfs_mfinal_met}.}
    \label{fig:pnlfs_mw} 
\end{figure}

While we have shown that the general shape of the PNLF is qualitatively in agreement with observations, the question remains whether the absolute number of PNe found at a given \OIII{} magnitude is comparable with observations, that is the normalization of the PNLF. To not only compare this for the brightest PNe, but obtain a result reaching faint PNe as well, we decided to use the PNLF measured for PNe within \SI{2}{\kilo\parsec} of the Sun by \citet{frew08}. That PNLF is estimated to be statistically complete up to six orders of magnitude below the bright end cutoff and can be seen in \cref{fig:pnlfs_mw}.

From the simulation, we selected a disk galaxy from Magneticum Box4 (uhr) with a Milky Way-like stellar mass ($M_* = \SI{3e10}{\Msun}$) to determine an analogous PNLF in the outer regions of the disk. For this, we only considered the stellar particles within a radius of \SI{2}{\kilo\parsec} from a point in the disk \SI{8}{\kilo\parsec} away from the galaxy center. The average stellar age in that region is $\langle t_\mathrm{age} \rangle = \SI{5.6}{\giga\year}$ and the average metallicity $\langle Z \rangle = 0.027$, which is overall similar to the solar neighborhood. The resulting PNLF from PICS using the metallicity-dependent model sequence is very similar to that from \citet{frew08}, lying only around \SI{0.2}{\dex} higher than the observed PNLF at the dim end and dropping off slightly earlier at the bright end (\cref{fig:pnlfs_mw}). As the simulated galaxy is of course not an exact Milky Way-analog, it is expected to also have differences with respect to its PNLF. However, the fact that the overall normalization and absolute number of PNe down to six orders of magnitude below the bright end are in agreement between the observations and simulations shows that the PNLF normalization obtained from the basic metallicity-dependent PICS model sequence is reasonable. In addition, it also shows that the assumption of PNe sometimes being optically thin is necessary to obtain the correct normalization since optically thick PNe lead to more particles being ionized and thus more \OIII{} emission. It should be noted that a perfect comparison is not only difficult because of the differences between the simulated galaxy and the Milky Way, but also because of the circumstellar extinction not being taken into account for the Milky Way PNe.

Finally, we also determined the PNLF of the inner stellar particles within \SI{2}{\kilo\parsec} of the galaxy center (red line in \cref{fig:pnlfs_mw}). Due to the larger density of stars in the center, there are significantly more PNe in this region. Additionally, the shape of the PNLF is different than that of the disk, which is likely due to the bulge of the simulated galaxy being younger and more metal-rich than the disk (unlike the Milky Way bulge). This difference in stellar population affects the shape of the PNLF, as already observed in previous studies \citep[e.g.,][]{reid&parker10:III,valenzuela+19,bhattacharya+21:III} and will be further investigated in a future study.

\section{Summary and Conclusion}
\label{sec:conclusion}

In this work, we have introduced a novel method of integrating PNe in cosmological simulations through post-processing, called PICS (PNe In Cosmological Simulations), and have presented first results from it. The method consists of a sequence of models that define the stellar lifetime function, the initial-to-final mass relation, the post-AGB stellar evolutionary tracks, and the nebular model. We employ the models of \citet{miller_bertolami16} and \citet{valenzuela+19} and use the lifetime function from \citet{padovani&matteucci93} and \citet{renzini&buzzoni86}, which has been used independently of metallicity in the past, to compare the effect that metallicity-dependent models have: We show that a reasonable bright end of the PNLF of old metal-rich stellar populations can only be obtained when taking into account that metal-poor stars reach the PN phase more quickly than metal-rich stars of the same mass.

We also demonstrate that the metallicity-dependent model sequence of PICS is capable of producing the same shape and normalization (i.e., the same absolute number of PNe) of the PNLF as is observed in the Milky Way disk around the Sun, down to at least six magnitudes below the bright end of the PNLF. As a result, PICS is shown to produce reasonably accurate populations of PNe and PNLFs for different underlying stellar populations. This work therefore lays the foundation for a more detailed analysis of the drivers behind the properties of the PNLF as well as for studies of the PNLF for galaxies across a diverse range of morphologies and formation histories, promising to significantly improve our understanding of the PNLF.

\section*{Acknowledgments}
LMV acknowledges support by the German Academic Scholarship Foundation (Studienstiftung des deutschen Volkes) and the Marianne-Plehn-Program of the Elite Network of Bavaria.

\bibliographystyle{mnras}
\bibliography{bib}

\end{document}